\input harvmac
\def\ZZ{\hbox{Z\kern-.4emZ}}
\def\RR{\hbox{R\kern-.6emR}}
\def\ZZs{\hbox{\zfonteight Z\kern-.4emZ}}

\lref\StromingerSH{
  A.~Strominger and C.~Vafa,
  ``Microscopic Origin of the Bekenstein-Hawking Entropy,''
  Phys.\ Lett.\  B {\bf 379}, 99 (1996)
  [arXiv:hep-th/9601029].
}

\lref\CastroMS{
  A.~Castro, D.~Grumiller, F.~Larsen and R.~McNees,
  ``Holographic Description of AdS$_2$ Black Holes,''
  JHEP {\bf 0811}, 052 (2008)
  [arXiv:0809.4264 [hep-th]].
}

\lref\HartmanDQ{
  T.~Hartman and A.~Strominger,
  ``Central Charge for AdS$_2$ Quantum Gravity,''
  JHEP {\bf 0904}, 026 (2009)
  [arXiv:0803.3621 [hep-th]].
}

\lref\GuicaMU{
  M.~Guica, T.~Hartman, W.~Song and A.~Strominger,
  ``The Kerr/CFT Correspondence,''
  arXiv:0809.4266 [hep-th].
}

\lref\SenQY{
  A.~Sen,
  ``Black Hole Entropy Function, Attractors and Precision Counting of
  Microstates,''
  Gen.\ Rel.\ Grav.\  {\bf 40}, 2249 (2008)
  [arXiv:0708.1270 [hep-th]].
}

\lref\BardeenPX{
  J.~M.~Bardeen and G.~T.~Horowitz,
  ``The extreme Kerr throat geometry: A vacuum analog of AdS(2) x S(2),''
  Phys.\ Rev.\  D {\bf 60}, 104030 (1999)
  [arXiv:hep-th/9905099].
}

\lref\BalasubramanianKQ{
  V.~Balasubramanian, A.~Naqvi and J.~Simon,
  ``A multi-boundary AdS orbifold and DLCQ holography: A universal  holographic
  description of extremal black hole horizons,''
  JHEP {\bf 0408}, 023 (2004)
  [arXiv:hep-th/0311237].
}
\lref\BalasubramanianBG{
  V.~Balasubramanian, J.~de Boer, M.~M.~Sheikh-Jabbari and J.~Simon,
  ``What is a chiral 2d CFT? And what does it have to do with extremal black
  holes?,''
  arXiv:0906.3272 [hep-th].
}

\lref\BarnichBF{
  G.~Barnich and G.~Compere,
  ``Surface charge algebra in gauge theories and thermodynamic integrability,''
  J.\ Math.\ Phys.\  {\bf 49}, 042901 (2008)
  [arXiv:0708.2378 [gr-qc]].
}
\lref\CompereAZ{
  G.~Compere,
  ``Symmetries and conservation laws in Lagrangian gauge theories with
  applications to the mechanics of black holes and to gravity in three
  dimensions,''
  arXiv:0708.3153 [hep-th].
}
\lref\BarnichKQ{
  G.~Barnich and G.~Compere,
  ``Conserved charges and thermodynamics of the spinning Goedel black hole,''
  Phys.\ Rev.\ Lett.\  {\bf 95}, 031302 (2005)
  [arXiv:hep-th/0501102].
}
\lref\BanadosDA{
  M.~Banados, G.~Barnich, G.~Compere and A.~Gomberoff,
  ``Three dimensional origin of Goedel spacetimes and black holes,''
  Phys.\ Rev.\  D {\bf 73}, 044006 (2006)
  [arXiv:hep-th/0512105].
}

\lref\EmparanEN{
  R.~Emparan and A.~Maccarrone,
  ``Statistical Description of Rotating Kaluza-Klein Black Holes,''
  Phys.\ Rev.\  D {\bf 75}, 084006 (2007)
  [arXiv:hep-th/0701150].
}

\lref\DiasNJ{
  O.~J.~C.~Dias, R.~Emparan and A.~Maccarrone,
  ``Microscopic Theory of Black Hole Superradiance,''
  Phys.\ Rev.\  D {\bf 77}, 064018 (2008)
  [arXiv:0712.0791 [hep-th]].
}

\lref\BardeenPX{
  J.~M.~Bardeen and G.~T.~Horowitz,
  ``The extreme Kerr throat geometry: A vacuum analog of AdS(2) x S(2),''
  Phys.\ Rev.\  D {\bf 60}, 104030 (1999)
  [arXiv:hep-th/9905099].
}

\lref\BalasubramanianRE{
  V.~Balasubramanian and P.~Kraus,
  ``A stress tensor for anti-de Sitter gravity,''
  Commun.\ Math.\ Phys.\  {\bf 208}, 413 (1999)
  [arXiv:hep-th/9902121].
}

\lref\SkenderisWP{
  K.~Skenderis,
  ``Lecture notes on holographic renormalization,''
  Class.\ Quant.\ Grav.\  {\bf 19}, 5849 (2002)
  [arXiv:hep-th/0209067].
}

\lref\AmselEV{
  A.~J.~Amsel, G.~T.~Horowitz, D.~Marolf and M.~M.~Roberts,
  ``No Dynamics in the Extremal Kerr Throat,''
  arXiv:0906.2376 [hep-th].
}
\lref\DiasEX{
  O.~J.~C.~Dias, H.~S.~Reall and J.~E.~Santos,
  ``Kerr-CFT and gravitational perturbations,''
  arXiv:0906.2380 [hep-th].
}

\lref\MaldacenaBW{
  J.~M.~Maldacena and A.~Strominger,
  ``AdS(3) black holes and a stringy exclusion principle,''
  JHEP {\bf 9812}, 005 (1998)
  [arXiv:hep-th/9804085].
}

\lref\CveticUW{
  M.~Cvetic and F.~Larsen,
  ``General rotating black holes in string theory: Greybody factors and  event
  horizons,''
  Phys.\ Rev.\  D {\bf 56}, 4994 (1997)
  [arXiv:hep-th/9705192].
}

\lref\CveticXV{
  M.~Cvetic and F.~Larsen,
  ``Greybody factors for rotating black holes in four dimensions,''
  Nucl.\ Phys.\  B {\bf 506}, 107 (1997)
  [arXiv:hep-th/9706071].
}

\lref\CveticVP{
  M.~Cvetic and F.~Larsen,
  ``Black hole horizons and the thermodynamics of strings,''
  Nucl.\ Phys.\ Proc.\ Suppl.\  {\bf 62}, 443 (1998)
  [Nucl.\ Phys.\ Proc.\ Suppl.\  {\bf 68}, 55 (1998)]
  [arXiv:hep-th/9708090].
}
\lref\CveticAP{
  M.~Cvetic and F.~Larsen,
  ``Greybody factors for black holes in four dimensions: Particles with
  spin,''
  Phys.\ Rev.\  D {\bf 57}, 6297 (1998)
  [arXiv:hep-th/9712118].
}

\lref\LarsenGE{
  F.~Larsen,
  ``A string model of black hole microstates,''
  Phys.\ Rev.\  D {\bf 56}, 1005 (1997)
  [arXiv:hep-th/9702153].
}

\lref\DijkgraafFQ{
  R.~Dijkgraaf, J.~M.~Maldacena, G.~W.~Moore and E.~P.~Verlinde,
  ``A black hole farey tail,''
  arXiv:hep-th/0005003.
}

\lref\KrausVZ{
  P.~Kraus and F.~Larsen,
  ``Microscopic Black Hole Entropy in Theories with Higher Derivatives,''
  JHEP {\bf 0509}, 034 (2005)
  [arXiv:hep-th/0506176].
}

\lref\LarsenBU{
  F.~Larsen,
  ``Anti-de Sitter spaces and nonextreme black holes,''
  arXiv:hep-th/9806071.
}

\lref\CveticXV{
  M.~Cvetic and F.~Larsen,
  ``Greybody factors for rotating black holes in four dimensions,''
  Nucl.\ Phys.\  B {\bf 506}, 107 (1997)
  [arXiv:hep-th/9706071].
}

\lref\MaldacenaDS{
  J.~M.~Maldacena and L.~Susskind,
  ``D-branes and Fat Black Holes,''
  Nucl.\ Phys.\  B {\bf 475}, 679 (1996)
  [arXiv:hep-th/9604042].
}
\lref\KunduriVF{
  H.~K.~Kunduri, J.~Lucietti and H.~S.~Reall,
  ``Near-horizon symmetries of extremal black holes,''
  Class.\ Quant.\ Grav.\  {\bf 24}, 4169 (2007)
  [arXiv:0705.4214 [hep-th]].
}

\lref\CveticUW{
  M.~Cvetic and F.~Larsen,
  ``General rotating black holes in string theory: Greybody factors and  event
  horizons,''
  Phys.\ Rev.\  D {\bf 56}, 4994 (1997)
  [arXiv:hep-th/9705192].
}

\lref\CveticXV{
  M.~Cvetic and F.~Larsen,
  ``Greybody factors for rotating black holes in four dimensions,''
  Nucl.\ Phys.\  B {\bf 506}, 107 (1997)
  [arXiv:hep-th/9706071].
}

\lref\CveticVP{
  M.~Cvetic and F.~Larsen,
  ``Black hole horizons and the thermodynamics of strings,''
  Nucl.\ Phys.\ Proc.\ Suppl.\  {\bf 62}, 443 (1998)
  [Nucl.\ Phys.\ Proc.\ Suppl.\  {\bf 68}, 55 (1998)]
  [arXiv:hep-th/9708090].
}

\lref\LarsenGE{
  F.~Larsen,
  ``A string model of black hole microstates,''
  Phys.\ Rev.\  D {\bf 56}, 1005 (1997)
  [arXiv:hep-th/9702153].
}

\lref\StromingerYG{
  A.~Strominger,
  ``AdS(2) quantum gravity and string theory,''
  JHEP {\bf 9901}, 007 (1999)
  [arXiv:hep-th/9809027].
}
\lref\ChoFZ{
  J.~H.~Cho, T.~Lee and G.~W.~Semenoff,
  ``Two dimensional anti-de Sitter space and discrete light cone
  quantization,''
  Phys.\ Lett.\  B {\bf 468}, 52 (1999)
  [arXiv:hep-th/9906078].
}

\lref\AzeyanagiBJ{
  T.~Azeyanagi, T.~Nishioka and T.~Takayanagi,
  ``Near Extremal Black Hole Entropy as Entanglement Entropy via AdS2/CFT1,''
  Phys.\ Rev.\  D {\bf 77}, 064005 (2008)
  [arXiv:0710.2956 [hep-th]].
}

\lref\GuptaKI{
  R.~K.~Gupta and A.~Sen,
  ``Ads(3)/CFT(2) to Ads(2)/CFT(1),''
  JHEP {\bf 0904}, 034 (2009)
  [arXiv:0806.0053 [hep-th]].
}

\lref\MaldacenaUZ{
  J.~M.~Maldacena, J.~Michelson and A.~Strominger,
  ``Anti-de Sitter fragmentation,''
  JHEP {\bf 9902}, 011 (1999)
  [arXiv:hep-th/9812073].
}

\lref\SenVZ{
  A.~Sen,
  ``Arithmetic of Quantum Entropy Function,''
  arXiv:0903.1477 [hep-th].
}

\lref\SenVM{
  A.~Sen,
  ``Quantum Entropy Function from AdS(2)/CFT(1) Correspondence,'' 
   arXiv:0809.3304 [hep-th].
}

\lref\HollandsYA{
  S.~Hollands, A.~Ishibashi and D.~Marolf,
  ``Counter-term charges generate bulk symmetries,''
  Phys.\ Rev.\  D {\bf 72}, 104025 (2005)
  [arXiv:hep-th/0503105].
}

\lref\MatsuoSJ{
  Y.~Matsuo, T.~Tsukioka and C.~M.~Yoo,
  ``Another Realization of Kerr/CFT Correspondence,''
  arXiv:0907.0303 [hep-th].
}

\lref\MatsuoPG{
  Y.~Matsuo, T.~Tsukioka and C.~M.~Yoo,
  ``Yet Another Realization of Kerr/CFT Correspondence,''
  arXiv:0907.4272 [hep-th].
}
\lref\BanerjeeUK{
  N.~Banerjee, I.~Mandal and A.~Sen,
  ``Black Hole Hair Removal,''
  arXiv:0901.0359 [hep-th].
}
\lref\SenVZ{
  A.~Sen,
  ``Arithmetic of Quantum Entropy Function,''
  arXiv:0903.1477 [hep-th].
}
\lref\BredbergPV{
  I.~Bredberg, T.~Hartman, W.~Song and A.~Strominger,
  ``Black Hole Superradiance From Kerr/CFT,''
  arXiv:0907.3477 [hep-th].
}

\lref\RasmussenIX{
  J.~Rasmussen,
  ``Isometry-preserving boundary conditions in the Kerr/CFT correspondence,''
  arXiv:0908.0184 [hep-th].
}

\lref\AmselPU{
  A.~J.~Amsel, D.~Marolf and M.~M.~Roberts,
  ``On the Stress Tensor of Kerr/CFT,''
  arXiv:0907.5023 [hep-th].
}

\lref\BarnichJY{
  G.~Barnich and F.~Brandt,
  ``Covariant theory of asymptotic symmetries, conservation laws and  central
  charges,''
  Nucl.\ Phys.\  B {\bf 633}, 3 (2002)
  [arXiv:hep-th/0111246].
}

\lref\CompereIN{
  G.~Compere and S.~Detournay,
  ``Centrally extended symmetry algebra of asymptotically Goedel spacetimes,''
  JHEP {\bf 0703}, 098 (2007)
  [arXiv:hep-th/0701039].
}

\Title{
}
{\vbox{\centerline{Near Extremal Kerr Entropy from AdS$_2$}
\medskip
\centerline{Quantum Gravity}}}
\medskip
\centerline{\it
Alejandra Castro${}^{\spadesuit,\dagger}$\foot{acastro@physics.mcgill.ca} and Finn Larsen${}^{\dagger}$\foot{larsenf@umich.edu}
}
\bigskip
\centerline{${}^\spadesuit$Physics Department, McGill University, Montreal, QC H3A 2T8, Canada.}
\smallskip
\centerline{${}^\dagger$Michigan Center for Theoretical Physics, Ann Arbor,
MI 48109, USA.}
\smallskip

\vglue .3cm
\bigskip\bigskip\bigskip
\centerline{\bf Abstract}
\noindent
We analyze the asymptotic symmetries of near extremal Kerr black holes
in four dimensions using the AdS$_2$/CFT$_1$ correspondence. 
We find a Virasoro algebra with central charge $c_R=12J$ that is independent 
from the Virasoro algebra (with the same central charge) that acts on the degenerate 
ground state. The energy of the excitations is computed as well, and we can use 
Cardy's formula to determine the near extremal entropy. Our result is consistent with the 
Bekenstein-Hawking area law for near extremal Kerr black holes. 

\Date{}

\newsec{Introduction}
The understanding of black hole entropy in string theory generally involves
the detailed study of specific microscopic configurations. A general 
shortcoming of this approach is that it obscures the presumed universal
physics underlying black hole thermodynamics. To address this situation there has 
recently been renewed interest in more generic black holes, including 
those that are of interest in the astrophysical setting. 
An important system for this research direction is the extreme Kerr black 
hole \refs{\GuicaMU,\BredbergPV}. 
The near horizon geometry of all extreme black 
holes has an AdS$_2$ component \KunduriVF\ and, for extreme Kerr, this feature has 
been combined 
with diffeomorphism invariance to give what may be a more robust understanding of black 
hole thermodynamics. 

In this paper we generalize these results further, by accounting for the entropy of near 
extreme Kerr black holes. This generalization is interesting because it introduces 
new theoretical features that are expected to be present also for generic black holes. 
Additionally, astrophysical applications often involve Kerr that are nearly extreme,
but not fully extreme. 

The conjectured description of extremal Kerr black hole is in terms of a chiral CFT
with central charge \GuicaMU\
\eqn\aa{
c_L = 12J~.
}
The internal structure of the black hole is encoded in the ground state degeneracy of this 
chiral CFT. In our description, the excitations responsible for the leading departure from 
extremality are described by the {\it same} CFT, except that we must focus on
the {\it opposite} chirality. We compute the boundary current of this
sector and derive the corresponding central charge which is
\eqn\ab{
c_R = 12J~.
}
The two chiral sectors thus have the same central charge. However, they should not be
confused since they are quite different in other aspects.

Our result for the boundary charges determines the relation between excitation
energy and CFT level $h_R$ for the right movers. We find that, in the right sector, 
there is no ground state degeneracy in the classical limit. The entropy is carried by 
genuine excitations and counted by Cardy's formula
\eqn\ac{
S_R = 2\pi \sqrt{ c_R h_R \over 6}~.
}
The result we find in the CFT agrees with the Bekenstein-Hawking area law
for near-extreme Kerr black holes. 

In the original Kerr/CFT \GuicaMU, the near horizon Kerr geometry (NHEK) \BardeenPX\ 
is represented as a warped U(1) fibration over AdS$_2$. The L Virasoro algebra emerges 
from the diffeomorphisms acting non-trivially on the asymptotic space of the near horizon 
geometry, following \refs{\BarnichJY,\BarnichBF,\CompereAZ,\BarnichKQ,\CompereIN}. 
In this construction only a $U(1)$ subgroup of the Virasoro algebra commutes 
with the $SL(2,\RR)$ isometry of the AdS$_2$ geometry. Recently, a similar strategy 
was pursued to study the finite temperature effects for the Kerr black 
hole \refs{\MatsuoSJ,\MatsuoPG}.

The near extreme generalization we present is much closer in spirit to conventional 
AdS/CFT correspondence. We follow \CastroMS\ 
(which in turn generalized \HartmanDQ) 
and exploit a complementary set of diffeomorphisms which preserve the asymptotic AdS$_2$ 
geometry. This yields the R Virasoro algebra, with an $SL(2,\RR)$ subgroup that can be 
identified with the isometry group. Excitations to finite level of this Virasoro algebra gives
the entropy above extremality. 

Our implementation of the AdS/CFT correspondence is closely related to Sen's 
quantum entropy function \refs{\SenQY,\BanerjeeUK,\SenVZ}. 
For example, the boundary mass term we introduce to regulate the on-shell action is 
reminiscent of the Wilson line that one must introduce in the quantum entropy function. 
It would of course be interesting to make such relations precise. However, our focus in 
this paper is to generalize the constructions to finite temperature. We anticipate that such 
a generalization applies to the quantum entropy function as well. 

The interplay between the two chiral sectors of the near extreme Kerr/CFT can be 
illuminated by contemplating a more general theory describing black holes far from 
extremality, inspired by \refs{\LarsenGE,\CveticUW,\CveticXV,\CveticVP}
and \refs{\EmparanEN,\DiasNJ}\foot{We describe this 
model a bit further in the final discussion.}. 
Suppose such a theory has the structure of a string theory, {\it i.e.} two chiral 
sectors coupled by some level matching condition that operates on the zero-modes alone. 
One sector (R) has excitations with the ability to carry angular momentum, the other (L)
has quanta that cannot carry angular momentum. The Kerr black hole rotates, so in this case
a macroscopic number of R quanta with angular momentum are excited. 
In the extreme limit the R sector becomes a condensate and {\it only} the quanta that carry 
angular momentum are excited. In this case the classical entropy derives from the degeneracy of 
the L sector which, due to level matching, must be excited as well. The entropy of
supersymmetric black holes similarly derives from excitations in the L sector but, unlike extremal
Kerr, the R sector is in its supersymmetric ground state. The entropy of non-supersymmetric 
extreme black holes (like the $D0-D6$ black holes) are the opposite, it comes entirely 
from excitations in the R sector, and so breaks supersymmetry completely. 

According to this model, the near extreme Kerr black hole receive most of their entropy from the
ground state entropy in the L sector, but non-extremality involves exciting the R sector beyond 
those quanta that carry the angular momentum. It is the R sector above the condensate that
we probe in this paper. 

We derive our results directly from the 2D perspective. However, it is also interesting to 
analyze AdS$_2$/CFT$_1$ correspondence by embedding into the more familiar 
AdS$_3$/CFT$_2$ correspondence
\refs{\StromingerYG,\ChoFZ,\AzeyanagiBJ,\BalasubramanianKQ,\CastroMS,\GuptaKI,\BalasubramanianBG}. In this construction the states responsible for the ground state
entropy are invariant under the $SL(2,\RR)$ that appears as an isometry. The excitations
above the ground state are quite different: they have a dispersion relation quadratic in the 
energy, and they do transform under the $SL(2,\RR)$. It is the physics of those excitations 
we focus on.

This paper is organized as follows. 
In section 2 we present a near horizon limit of the Kerr black hole that maintains energy 
above extremality. A component of this near horizon region is asymptotically (and locally) 
AdS$_2$. We also show how to present near extreme Kerr entropy as a sum of ground state
entropy and excitations described by Cardy's formula. 
In sections 3 through 6 we develop the properties of the near horizon Kerr from the perspective of AdS$_2$ quantum gravity.  In section 3 we construct the effective 2D theory by dimensional reduction of the four dimensional solution and classify the solutions to this theory. In section 4 we renormalize the action by adding boundary counterterms, ensuring 
a well-defined variational principle and finite conserved charges.
In section 5 we determine the diffeomorphisms that preserve the asymptotic AdS$_2$
behavior, as well as other gauge conditions. This will follow the procedure
in \CastroMS\ very closely, albeit with an effective 2D action that is slightly different from the
one analyzed \CastroMS. In section 6 we construct the conserved Noether charges associated to the asymptotic symmetries. We find that the charge associated to diffeomorphisms is a non trivial combination of the boundary stress tensor and the $U(1)$ current. In section 7 we employ our findings from previous sections to investigate the properties of the near extremal black hole.
Comparison with the area law gives a perfect match. 
Finally, in section 8, we conclude by discussing some of the open problems and known clues that point to a more 
general CFT description of all 4D black holes, including the elusive Schwarzchild 
black holes. 

\newsec{The Near Extreme Kerr Limit}

In this section we isolate the near horizon geometry of the Kerr black hole in a 
manner that maintains some excitation energy above the extremal limit. We discuss
the extreme and near extreme Kerr entropy.

\subsec{Near Horizon Geometry}
The general Kerr solution is given by
\eqn\xa{\eqalign{ds^2=&-{\Sigma\Delta\over (r^2+a^2)^2-\Delta a^2\sin^2\theta}dt'^2+{\Sigma}\left[{dr^2\over\Delta}+d\theta^2\right]\cr & +{\sin^2\theta\over\Sigma}((r^2+a^2)^2-\Delta a^2\sin^2\theta)\left[d\phi'-{2a\mu r\over (r^2+a^2)^2-\Delta a^2\sin^2\theta}dt'\right]^2~,}}
with
\eqn\xab{\Delta=(r-r_-)(r-r_+)~,\quad r_\pm = \mu\pm \sqrt{\mu^2-a^2~,}}
and
\eqn\xac{\Sigma=r^2+a^2\cos^2\theta~.}
In our notation $\mu = G_4M$ and $a=J/M$ are length scales, while $J$ is dimensionless. The 
near horizon region is isolated by introducing the coordinates
\eqn\xad{r={1\over2}(r_++r_-)+\lambda U~,\quad t'={t\over \lambda}~,\quad \phi' = \phi+{t\over \lambda (r_++r_-)}~.}
The dimensionless scaling parameter $\lambda\to 0$ in the near horizon limit, with $t, U, \theta, \phi$
fixed. 

For extremal black  holes we take the near horizon scaling limit with $r_+ =r_-$ from the outset.
The near horizon geometry becomes 
\eqn\xae{\eqalign{ds^2={1+\cos^2\theta\over2}\left[-{U^2\over \ell^2}dt^2+{\ell^2\over U^2}dU^2+
\ell^2 d\theta^2\right]+\ell^2{2\sin^2\theta\over1+\cos^2\theta}\left(d\phi+{U\over\ell^2}dt\right)^2~,}}
where we defined
\eqn\xaf{\ell^2 \equiv {1\over 2}(r_++r_-)^2=2\mu^2~.}
The metric \xae\ is known as Near Horizon of Extreme Kerr (NHEK) \BardeenPX. The components 
described by the $(t,U)$ coordinates in the square bracket of \xae\ is an AdS$_2$ geometry, 
with radius $\ell$ given in \xaf. 

The near horizon limit is easily modified to maintain some energy above extremality
in the limit. We need to take the limit while tuning the black hole parameters
such that the scale $\epsilon$ defined through 
\eqn\xb{{1\over 2}{(r_+-r_-)}=\sqrt{\mu^2-a^2}\equiv \epsilon\lambda~,}
is kept fixed as $\lambda\to 0$. The radius $\ell$ given in \xaf\ is also fixed. Using
\eqn\xba{\eqalign{{dr^2\over \Delta}&={dU^2\over U^2-\epsilon^2}~,\cr \Sigma&= {\ell^2\over 2}(1+\cos^2\theta)+{\cal O}(\lambda)~,\cr  (r^2+a^2)^2-\Delta a^2\sin^2\theta&=\ell^4 [1 + {\sqrt{8}\over\ell}\lambda U  +{\cal O}(\lambda^2)] ~,}}
the resulting line element reads
\eqn\xbc{\eqalign{ds^2=&{1+\cos^2\theta\over2}\left[-{U^2-\epsilon^2\over \ell^2}dt^2+{\ell^2\over U^2-\epsilon^2}dU^2+\ell^2d\theta^2\right]+\ell^2{2\sin^2\theta\over1+\cos^2\theta}\left(d\phi+{U\over\ell^2}dt\right)^2~.}}
This near extreme horizon geometry modifies the original NHEK geometry \xae\ only in 
the AdS$_2$ component. In fact, the $(t,U)$ term in the square brackets remains 
locally AdS$_2$ with radius of curvature $\ell$ but the global structure is modified into a black 
hole geometry with horizon located at  $U=\epsilon$. We interpret this deformation as an 
excitation above the extreme Kerr. The near extreme limit described here is the
same considered in \refs{\AmselEV,\DiasEX} (and a Kerr analogue 
of  ``Limit 2'' in \MaldacenaUZ).


\subsec{Phenomenology of Kerr Thermodynamics}
It is instructive to consider Kerr thermodynamics in the near extreme limit. 

When manipulating thermodynamic formulae it is useful to introduce the
Planck mass $M_P$ and Planck length $l_P$ through 
\eqn\ja{
G_4 = M^{-2}_P= l^2_P ~.
}
In this notation, we are interested in near extremal Kerr black holes for which
\eqn\jb{
M = \sqrt{J} M_P + E~,
}
with the excitation energy 
$E\ll \sqrt{J}M_P$. In this regime an important scale is the AdS$_2$ scale \xaf\ 
\eqn\jc{
\ell = {1\over\sqrt{2}}(r_+ + r_-) = \sqrt{2} Ml_P^2 = \sqrt{2J}l_P + \ldots~.
}
We can recast the precise expansion parameter here and in similar expressions below as 
$E\ell/J \ll 1$.

The black hole entropy for a general Kerr black hole is
\eqn\ja{\eqalign{
S & ={A\over 4G_4} =  2\pi (M^2 l^2_P + \sqrt{M^4 l^4_P - J^2})~.\cr
}}
Expanding in the near extremal limit \jb, the entropy becomes
\eqn\jd{\eqalign{
S & = 2\pi\left( J + \sqrt{4J^{3/2}E l_P} \right) + {\cal O} (E \ell)~.
}}
The expression takes the suggestive form
\eqn\je{
S = 2\pi \left( {c_L\over 12} + \sqrt{c_Rh_R\over 6}\right) + \ldots~,
}
with central charges
\eqn\jf{
c_L = c_R = 12J~,
}
and weight
\eqn\jg{
h_R  = 2E\sqrt{J}l_P~.
}
In summary: in the extremal case $E=0$, the entropy is a ground state entropy, 
corresponding to Cardy's formula with $h_L={c\over 24}$. On the other hand, the 
departure from the extremal entropy is given by a more conventional Cardy formula.

In our decoupling limit we parametrize the excitation energy by the nonextremality 
parameter $\epsilon$ (introduced in \xb). It is related to the excitation energy 
$E$ (introduced in \jb) through
\eqn\ji{
\epsilon \lambda = \sqrt{\mu^2 - a^2} = {1\over M}\sqrt{M^2 l^4_P - J^2}
= {1\over\sqrt{JM^2_P}} \sqrt{4J^{3/2} E l_P}
= \sqrt{ 4\sqrt{J}El_P^3} ~.
}
The weight  \jg\ assigned to the excitation energy becomes
\eqn\jj{
h_R= 2E\sqrt{J}l_P= {\epsilon^2\lambda^2\over 2l_P^2}~. 
}
The expected level is small because typical (comoving) energies in the near horizon 
are small compared to typical asymptotic energies. According to \xad\ we have 
$\partial_t=\lambda\partial_{t'}$. However, since the level \jj\ is of order
${\cal O}(\lambda^2)$, the excitation energy will be parametrically small even 
compared to typical energies in AdS$_2$. The ``non-relativisitic'' form \jj\ of
the energy is wellknown from lightcone physics. In the present context the relevant
excitations move at nearly the speed of light due to the rapid rotation of the black hole.

In order to ``understand'' the near extreme Kerr entropy \je\ we must account for the
two central charges \jf\ and the excitation level \jj. In the next four sections we develop
relevant results before returning to Kerr thermodynamics in section 7. 

\newsec{The 2D Perspective}
Our approach to Kerr dynamics focusses on the AdS$_2$ component of the
near horizon geometry \xbc. In this section we determine the 2D effective theory 
in which this geometry appears as a solution. 

\subsec{The Effective 2D Theory}
The 2D theory contains
a general 2D metric
\eqn\fea{ds^2_2={g}_{\mu\nu}dx^\mu dx^\nu~,}
with $\mu,\nu=1,2$, a gauge connection encoding the rotation
\eqn\feb{{\cal B}={\cal B}_\mu(x)dx^\mu~,}
and one scalar field $\psi$ that couples to the size of the two angular coordinates. 
In other words, we consider those 4D field configurations that take the form
\eqn\ffd{ds^2_4={1\over2}(1+\cos^2\theta)\left[ds^2_2+e^{-2\psi}\ell^2d\theta^2\right]
+e^{-2\psi}\ell^2\left({2\sin^2\theta\over1+\cos^2\theta
}\right)\left[d\phi+{\cal B}\right]^2~.}
In the following we raise and lower all indices using $g_{\mu\nu}$ 
and the covariant derivatives $\nabla_\mu$ are also formed using 
$g_{\mu\nu}$. The gauge field strength is denoted ${\cal G}=d{\cal B}$. 

The 4D action is simply the Einstein Hilbert action
\eqn\ea{
S_4 = {1\over 16\pi G_4}\int d^4 x \sqrt{-g_4}R^{(4)}~.
}
The class of 4D metrics we consider \ffd\ have Ricci scalar 
\eqn\ffe{\eqalign{R^{(4)}=&{2\over
1+\cos^2\theta}\left[{R^{(2)}}+e^{2\psi}{2\over\ell^2}-2\,e^{2\psi}\,{\nabla}^2e^{-2\psi}
+{1\over2}\,e^{4\psi}{\nabla}_{\mu}e^{-2\psi}\,{\nabla}^{\mu}e^{-2\psi}\right]\cr&
-2\left(1+\cos^2\theta\right)^{-3}\sin^2\theta\left[e^{-2\psi}\ell^2{\cal
G}^2+e^{2\psi}{2\over \ell^2}\right]~,}}
and determinant
\eqn\ff{\sqrt{-g_4}={\ell^2\over
2}e^{-2\psi}\sin\theta\left(1+\cos^2\theta\right)\sqrt{{-g}}~.}
Inserting these expressions into the 4D action \ea\ and integrating by parts 
using
\eqn\fga{\int_0^{2\pi} d\phi \int_{0}^\pi d\theta
\left(1+\cos^2\theta\right)^{-2}\sin^3\theta =2\pi
~,}
we find the 2D effective action
\eqn\fgb{S_{\rm Kerr}={\ell^2\over 4G_4}\int
d^2x\sqrt{{-g}}\left[e^{-2\psi}{R^{(2)}}+{1\over\ell^2}
+2{\nabla}_{\mu}e^{-\psi}\,{\nabla}^{\mu}e^{-\psi}-{\ell^2\over2}e^{-4\psi}{\cal
G}^2\right]~.}
We want to analyze this action. It is almost identical to the action considered in 
\CastroMS, but the dependence on the dilaton is different. 

\subsec{Solutions to the 2D Theory}
To get started, we determine some of the solutions to the 2D effective theory. 
For constant $\psi$, the equations of motion are
\eqn\fgc{\eqalign{{R^{(2)}}-e^{-2\psi}\ell^2{\cal G}^2&=0~,\cr
{1\over2}\left(e^{4\psi}{1\over\ell^2}-{\ell^2\over2}{\cal
G}^2\right){g}_{\mu\nu}+{\ell^2}\,{\cal G}_{\mu\rho}{\cal
G}_{\nu}^{~\rho}&=0~,\cr {\nabla}_\mu{\cal G^{\mu\nu}}&=0~.}}
Contracting the second equation in \fgc\ we get
\eqn\fgd{{\cal G}^2= -e^{4\psi}{2\over\ell^4}~,}
and the curvature is given by
\eqn\fge{{R^{(2)}}=-e^{2\psi}{2\over\ell^2}~.}
The solutions with constant $\psi$ are therefore locally AdS$_2$ with radius
\eqn\hel{\ell_{\rm AdS}=e^{-\psi}\ell~.}
We will loosely refer to $\ell$ as the AdS$_2$ radius, since we can take $\psi=0$ after 
solving the corresponding equation of motion.

The near horizon Kerr geometry \xbc\ should be a solution to this theory. 
Indeed, we can verify that this is the case with 
\eqn\fgz{\eqalign{
e^{-2\psi} &=1~,\cr
\ell^2&=2G_4J~.
}}
Also, inserting the equations of motion \fgd,\fge\ in the 2D action \fgb, we find that the on-shell bulk action vanishes, 
as expected since $R^{(4)}=0$ for the Kerr solution.

It is instructive to consider more general solutions. Without loss of generality 
we can choose a gauge where the metric takes the form
\eqn\hba{ds^2=e^{-2\psi}d\rho^2+\gamma_{tt}dt^2~,}
and the gauge field satisfies ${\cal B}_\rho=0$. The solutions to the equations of
motion in this case 
are\foot{The notation is related to that of \CastroMS\
through $f(t) = - h_1(t)$, $\ell=L/2$, $\gamma_{tt}=h_{tt}$.}
\eqn\hbc{\eqalign{
\gamma_{tt} &=-{1\over4}e^{-2\psi}\left(e^{\rho/\ell}-f(t)e^{-\rho/\ell}\right)^2~,\cr
{\cal B}_t
&={1\over2\ell}e^{\rho/\ell} \left(1- \sqrt{f(t)}e^{-\rho/\ell}\right)^2~.
}}
The $\rho$-independent term in ${\cal B}_t$ is arbitrary, a residual gauge
function. We chose that term so that ${\cal B}_t$ becomes a complete square, because 
this is the correct gauge for applications to black holes \SenVM. The 
function $f(t)$ is arbitrary (except that $f(t)\geq 0$ in the gauge we have presented 
the solution in). 

For any function $f(t)$, the asymptotic  behavior 
\eqn\hbf{\eqalign{
\gamma_{tt}^{(0)}&=-{1\over4}e^{-2\psi^{(0)}}e^{2\rho/\ell}~,\cr
{\cal B}_t^{(0)} &={1\over2\ell}e^{\rho/\ell}~,\cr  \psi^{(0)} & =
\hbox{constant}~,
}}
defines asymptotically AdS$_2$ configurations. We do not impose boundary
conditions on the other AdS$_2$ boundary. It may be an issue that the 
solutions \hbf\ are non-normalizable there \BalasubramanianBG. On the
other hand, in the Euclidean theory it is certainly consistent to focus on
solutions with one boundary \foot{We thank A. Sen for this argument.}. 

The effective 2D geometries 
appearing in extremal Kerr \xae\ and near-extremal Kerr \xbc\ are recovered
through the coordinate transformation
\eqn\hxa{
U = {\ell\over 2} e^{\rho/\ell}\left( 1 + {\epsilon^2\over\ell^2}e^{-2\rho/\ell}\right)~.
}
The value $f(t)=-1$ 
corresponds to global AdS$_2$, the value $f(t)=0$ corresponds to the extremal 
Kerr solution \xae, and
\eqn\hxb{
f(t) = {\epsilon^2\over\ell^2}~,
}
are the near extremal Kerr solutions \xbc. The terminology mimics the corresponding
situation for BTZ solutions in AdS$_3$. 

For later calculations, we record the extrinsic curvature at the boundary,
\eqn\hbe{K={1\over2}\gamma^{tt}n^\mu\partial_\mu\gamma_{tt}={1\over
\ell}e^{\psi}~,}
where the normal vector has component $n^{\rho}=\sqrt{g^{\rho\rho}}=e^{\psi}$.

\newsec{The Renormalized Boundary Action}
In this section we determine the renormalized boundary action of the
AdS$_2$ space. Our treatment follows the standard procedure for AdS/CFT 
correspondence in any dimension \refs{\BalasubramanianRE\SkenderisWP}, 
with the adaptation to AdS$_2$ developed in \refs{\SenVM,\CastroMS,\SenVZ}.

\subsec{Boundary Counterterms}
We first need to determine the boundary terms 
\eqn\hc{S_{\rm bndy} = S_{\rm GH}+S_{\rm counter}~,}
that give a well-defined variational principle. The Gibbons-Hawking term is given by
\eqn\hca{S_{\rm GH}={\ell^2\over 2G_4}\int
dt\sqrt{-\gamma}e^{-2\psi}K~.}
The counterterms take the local form 
\eqn\hcb{S_{\rm counter}={\ell^2\over 2G_4}\int
dt\sqrt{-\gamma}\left[\lambda e^{-\psi}+me^{-3\psi}{\cal B}_a{\cal B}^a\right]~,}
with $\lambda$ and $m$ constants to be determined. The power of
$e^{-\psi}$ in each term in \hcb\ is determined by the dependence on
$\psi$ of the metric and gauge field. Using the expressions from the
previous section we have $K\sim e^{\psi}$ and ${\cal B}_a{\cal
B}^a\sim e^{2\psi}$, then each term in \hca\ and \hcb\ scale as
$e^{-2\psi}$ and therefore the constants $\lambda$ and $m$ will only
depend on the scale $\ell$. The consistency of the second term in \hcb\ with gauge
invariance was established in \CastroMS\ for any gauge function that remains
finite at infinity. 

The variation of the full action, including boundary terms, takes the form
\eqn\hcc{\delta S=\int dt\sqrt{-\gamma}\left[\pi^{ab}\delta
\gamma_{ab}+\pi_\psi\delta\psi+\pi^a\delta {\cal B}_a\right]+\hbox{bulk
terms}~,}
where the $\pi$ are momenta that receive contributions from the variations of 
both the bulk and the boundary action. By explicit computation we find
\eqn\hcd{\eqalign{\pi^{tt}&={\ell^2\over 4
G_4}\left(\lambda e^{-\psi}\gamma^{tt}+me^{-3\psi}\gamma^{tt}{\cal B}_a{\cal
B}^a-2me^{-3\psi}{\cal B}^t{\cal B}^t\right)~, \cr
\pi^t&={\ell^2\over 4
G_4}\left(-2e^{-4\psi}\ell^2n_\mu{\cal G}^{\mu t}+4me^{-3\psi}{\cal
B}^t\right)~,\cr \pi_\psi&=-{\ell^2\over 2
G_4}\left(2e^{-2\psi}K+\lambda e^{-\psi}+3me^{-3\psi}{\cal B}_a{\cal
B}^a\right)~.}}
Using \hbe\ for the extrinsic curvature and \hbf\ for the asymptotic geometry we find
\eqn\hce{\eqalign{\pi^{tt}&={\ell^2\over 4
G_4}\left(\lambda +{m\over\ell^2}\right)e^{-\psi^{(0)}}\gamma^{tt}_{(0)}~,
\cr \pi^t&={\ell^2\over 4
G_4}\left(-1+2{m\over\ell}\right)e^{-3\psi^{(0)}}\gamma^{tt}_{(0)}e^{\rho/\ell}~,\cr
\pi_\psi&=-{\ell^2\over 2
G_4}\left(2{1\over\ell}+\lambda
-3{m\over\ell^2}\right)e^{-\psi^{(0)}}~.
}}
The vanishing of these three boundary momenta is ensured by the two conditions
\eqn\hcf{m={\ell\over2}~,\quad \lambda=-{1\over2\ell}~.}

Collecting results, we have determined the full action including boundary counterterms
\eqn\hcg{\eqalign{S=&{\ell^2\over 4 G_4}\int
d^2x\sqrt{{-g}}\left[e^{-2\psi}{R^{(2)}}+{1\over\ell^2}
+2{\nabla}_{\mu}e^{-\psi}\,{\nabla}^{\mu}e^{-\psi}-e^{-4\psi}{\ell^2\over2}{\cal
G}^2\right]\cr&+{\ell^2\over 2 G_4}\int
dt\sqrt{-\gamma}\left[e^{-2\psi}K-{1\over 2\ell}e^{-\psi}+{\ell\over2}e^{-3\psi}{\cal
B}_a{\cal B}^a\right]~.}}

\subsec{Boundary Currents}
As we will see in the following, there are special difficulties in two dimensions that
we have not yet addressed. Ignoring these for a moment, we proceed in the standard
manner and compute the response functions on the boundary
\eqn\hd{\eqalign{
T^{ab} &={2\over\sqrt{-\gamma}}{\delta S\over \delta \gamma_{ab}}~,\cr
J^{a}&={1\over\sqrt{-\gamma}}{\delta S\over \delta {\cal B}_{a}} ~.
}}
For the action \hcg, we find
\eqn\hda{\eqalign{T_{tt}&=-{\ell^2\over 4 G_4}\left(
{1\over\ell}e^{-\psi}\gamma_{tt}+\ell e^{-3\psi}{\cal B}_t{\cal
B}_t\right)~,\cr J_t&= {\ell^2\over 2
G_4}e^{-3\psi}\left(-e^{-\psi}\ell^2n^\mu{\cal G}_{\mu
t}+\ell{\cal B}_t\right)~.
}}

As an example, we evaluate these expressions for our AdS$_2$ black hole solution. 
For easy reference, we write the metric and the gauge field for the black hole which were previously given 
by \hbc\ with $f(t)$ the constant value \hxb,
\eqn\hdb{\eqalign{
\gamma_{tt} &= -{1\over 4} e^{-2\psi} e^{2\rho/\ell} \left(1 - {\epsilon^2\over\ell^2}e^{-2\rho/\ell}\right)^2~,\cr
{\cal B}_t & ={1\over 2\ell} e^{\rho/l}\left(   1- {\epsilon\over\ell} e^{-\rho/\ell}\right)^2~.
}}
As we noted earlier, the constant term in ${\cal B}_t$ is fixed in the black hole
context, by the condition that the corresponding Euclidean solution is regular
at the horizon. Now, inserting \hdb\ in \hda\ we find
\eqn\hde{\eqalign{
T_{tt}  &= 
{\ell\over 4G_4}e^{-3\psi}( {\epsilon\over\ell} e^{\rho/\ell} - {2\epsilon^2\over\ell^2} +\cdots)~,\cr
J_t  &
= {\ell^2\over 2G_4} e^{-3\psi}  \left( -{\epsilon\over\ell}+ {\epsilon^2\over\ell^2}e^{-\rho/\ell}\right)~.
}}
In each expression the leading divergence cancelled. Indeed, we designed the boundary 
counterterms to ensure this property. However, the remaining expression for the energy 
momentum tensor still diverges as $\rho\to\infty$. This behavior 
is special to two spacetime dimensions, and it is entirely unacceptable \foot{It is 
worth noting the response functions \hda\ would have been finite, if we had fixed 
the residual gauge so that there is no $\rho$-independent term in ${\cal B}_t$. 
The divergence in the energy momentum tensor can thus be removed at the expense 
of having a singular Euclidean gauge field at the horizon. This possibility may be 
relevant in some situations but we will not rely on that cancellation here.}. In 
the following sections we will carefully take into account some of the special features in
two spacetime dimensions  and find a more satisfactory result. 

\newsec{Boundary Conditions and Asymptotic Symmetries }
In the previous section we treated diffeomorphisms and gauge transformations as independent. 
However, Hartman and Strominger noticed that only a combination of these transformations are 
consistent with the gauge conditions \HartmanDQ. In this section we incorporate this feature. 

%

We have assumed that the metric is written in Fefferman-Graham form \hba, 
with $g_{\rho\rho}$ and $g_{\rho t}$ specified as a gauge conditions, along
with the asymptotic behavior \hbf. The diffeomorphisms that preserve
these conditions are those that satisfy
\eqn\hea{\delta_\epsilon g_{\rho\rho}=0~, \quad \delta_\epsilon
g_{\rho t}=0~, \quad \delta_\epsilon g_{tt}= 0\cdot
e^{2\rho/\ell}+\ldots~.}
General diffeomorphisms transforms the metric as
\eqn\he{\eqalign{\delta_{\epsilon}g_{\mu\nu}&=\nabla_\mu\epsilon_\nu+\nabla_\nu\epsilon_\mu\cr
&=g_{\nu\lambda}\partial_\mu\epsilon^\lambda+g_{\mu\lambda}\partial_\nu\epsilon^\lambda+
\epsilon^\lambda\partial_\lambda g_{\mu\nu}~.}}
Imposing the conditions \hea\ we find that the allowed coordinate transformations
are
\eqn\heb{\epsilon^\rho=-\ell\partial_t\xi(t)~,\quad
\epsilon^t=\xi(t)+{2\ell^2}\left(e^{2\rho/\ell} -f(t)\right)^{-1}\partial^2_t\xi(t)~,}
where $\xi(t)$ is an arbitrary function. We used the explicit solutions \hbc.
Under these allowed diffeomorphisms the boundary metric transforms as
\eqn\hec{\delta_\epsilon
\gamma_{tt}=-e^{-2\psi}(1 - f(t) e^{-2\rho/\ell})\left[ -f(t)\partial_t\xi
-{1\over2}\partial_t f(t)\,\xi+\ell^2\partial^3_t\xi\right]~.}

The next step is to consider the transformation of the gauge field ${\cal B}_\mu$ under the
allowed diffeomorphisms. A vector transforms according to
\eqn\hed{\eqalign{\delta_\epsilon {\cal
B}_\mu&=\epsilon^\lambda\nabla_\lambda {\cal B}_\mu +{\cal B}_\lambda
\nabla_\mu \epsilon^\lambda\cr 
&= \epsilon^\lambda\partial_\lambda
{\cal B}_\mu +{\cal B}_\lambda \partial_\mu \epsilon^\lambda~,
}}
under a general diffeomorphism. Considering just the allowed diffeomorphisms \heb, 
the transformation of the radial component becomes
\eqn\hef{\delta_\epsilon {\cal B}_\rho
=-2\left(1+\sqrt{f(t)}e^{-\rho/\ell}\right)^{-2}e^{-\rho/\ell}\partial_t^2\xi~.}
This clearly violates the gauge condition ${\cal B}_\rho=0$. 

To restore the gauge condition we compensate by a gauge transformation with gauge
function
\eqn\heg{\Lambda=-2\ell \,e^{-\rho/\ell}\left(1 + \sqrt{f(t)} e^{-\rho/\ell}\right)^{-1}\partial_t^2\xi
~,}
chosen so that the combined diffeomorphism and gauge transformation gives
\eqn\heh{\delta_{\epsilon +\Lambda}{\cal
B}_\rho=\delta_{\epsilon}{\cal B}_\rho+\partial_\rho\Lambda =0~.}
It is the combination of allowed diffeomorphisms \heb\ with the compensating gauge
transformation \heg\ that generate legitimate symmetries of the theory \HartmanDQ. 
The variation of the time component of the gauge field under the symmetry is
\eqn\hej{\delta_{\epsilon +\Lambda}{\cal
B}_t=-{e^{-\rho/\ell}\over \ell}\left[-f(t)\partial_t\xi - {1\over2}\partial_t
f(t) \,\xi+\ell^2\partial^3_t\xi\right]-{1\over\ell}\partial_t\left(\sqrt{f}\xi\right)~.}

The transformation of the stress tensor \hda\ under the combinations of 
diffeomorphisms and $U(1)$ transformations that constitute a symmetry
\eqn\hek{\eqalign{\delta_{\epsilon+\Lambda}T_{tt}&=T_{tt}\partial_t\xi
+\xi\partial_t T_{tt}+{\ell^3\over
2 G_4}e^{-3\psi}\partial_t^3\xi+{\cal O}(e^{-\rho/\ell})~, }}
and the transformation of the $U(1)$  current \hda\ under the same transformations is
\eqn\hex{\delta_{\epsilon+\Lambda}J_{t}=J_t\partial_t\xi+\xi\partial_tJ_t-{\ell^4\over
 G_4}e^{-3\psi}e^{-\rho/\ell}\partial_t^3\xi+{\cal O}(e^{-2\rho/\ell})~. }
%

%


The transformation laws are very interesting because they almost resemble the standard 
ones in CFT. In fact, if we introduce the central charge as
\eqn\hem{\delta_{\epsilon+\Lambda}T_{tt}=
T_{tt}\partial_t\xi
+\xi\partial_t T_{tt}-{c\over 12}\ell_{\rm AdS}\partial_t^3\xi~,}
where the AdS radius was defined in \hel,  gives
\eqn\hen{c={6\over G_4} \ell^2_{\rm AdS}~,}
up to a sign. The AdS$_2$ radius for near extreme Kerr \fgz\ then gives
\eqn\hep{c_{\rm  Kerr} = 12J~.}
This is exactly the value that was found in 
Kerr/CFT \GuicaMU\ using entirely different methods. Despite this apparent success,
essential concerns remain. Apart from the unacceptable sign, 
the stress-tensor $T_{tt}$ is not related in any obvious way to the combination of 
diffeomorphism and gauge transformation that is an actual symmetry of the system and it does not have the correct conformal weight. In 
the next section we develop a more systematic point of view that will determine the 
central charge unambiguously. 

\newsec{Conserved Charges}
In this section we construct the conserved currents that generate diffeomorphisms and gauge transformations. We also compute the central charge and $U(1)$ level associated to these generators.

\subsec{Noether Charge for Time Translation}
Under a diffeomorphism $x^{\mu}\to x^{\mu}+\epsilon^{\mu}$ the variation of the action is
\eqn\hdba{\delta_{\epsilon}S={1\over 2}\int dt \sqrt{-\gamma} \,T^{ab}\delta_\epsilon \gamma_{ab}+ \int dt \sqrt{-\gamma} J^{a}\delta_\epsilon {\cal B}_{a}+{\rm (e.o.m)}~,}
where 
 \eqn\hdc{\eqalign{\delta_\epsilon \gamma_{ab}&=\nabla_a \epsilon_b + \nabla_b \epsilon_a~,\cr
\delta_\epsilon {\cal B}_a&=\epsilon^b\nabla_b {\cal B}_a +{\cal B}_b\nabla_a \epsilon^b~.}}
To define the conserved charge associated to $\epsilon^a$, consider a regulator 
function $F$ such that in the neighborhood of the boundary we have $F=1$. Then 
define the variation of  a field $\Phi$ as 
\eqn\hdd{\delta_{\epsilon,F}\Phi\equiv\left( \delta_{F \epsilon}-F\delta_{\epsilon}\right)\Phi~.}
(For further details, see {\it e.g.} \HollandsYA ). Implementing this variation for the action we get
 \eqn\hdea{\eqalign{\delta_{\epsilon,F}S&=\int dt \sqrt{-\gamma} \,T^{ab}\epsilon_a\nabla_b F+ \int dt \sqrt{-\gamma} J^{a}{\cal B}_{b}\epsilon^b\nabla_a F\cr &= \sqrt{-\gamma} \gamma^{tt}
 (\,T_{tt}+J_{t}{\cal B}_{t})\epsilon^t -\int dt \,
 \nabla_b( \,T^{ba}\epsilon_a + J^{b}{\cal B}_{a}\epsilon^a) F ~,}}
where in the second line we integrated by parts and used $F=1$ near the
boundary. The last term vanishes due to Noether's theorem so the charge 
that generates infinitesimal temporal diffeomorphisms $\epsilon^t$ becomes
 \eqn\hdf{\eqalign{Q_\epsilon&= \sqrt{-\gamma}\gamma^{tt}\left(T_{tt}+J_t{\cal B}_t\right)~.}}
It is quite unfamiliar to find that the generator of translations involves a matter term, along
with the energy momentum tensor. The matter term does in fact appear in other contexts as 
well, but it is usually negligible because matter fields decay more rapidly than the metric field. 
The situation here is special since ${\cal B}_t\sim e^{\rho/\ell}$ increases 
rapidly as the boundary is approached. 
 
To make things explicit, we consider the black hole solution \hdb, for which the
expressions \hde\ give
\eqn\hdfa{
T_{tt} + {\cal B}_t J_t = {\ell\over 4G_4} e^{-3\psi} {\epsilon^2\over\ell^2} ~.
}
Recall that we previously found a divergence in the energy momentum tensor \hde. 
However, the additional matter term contains a divergence of its own, such that the sum is 
finite. 

The metric factors in \hdf\ redshift  $Q_\epsilon$ to zero in the limit $\rho\to\infty$. In this sense 
we will still assign vanishing energy to all excitations, as expected from general 
principles in two dimensions. (Some recent discussions 
are \refs{\AmselEV, \DiasEX, \BalasubramanianBG}). We will see in section 7.2 that
we nevertheless have a meaningful concept of energy. 
 
\subsec{The $U(1)$ Charge}

The electric charge is given by the variation of the action with respect to the field 
strength ${\cal G}$. We can implement this prescription by introducing an off-shell 
electric potential  $\phi_e$ as an overall proportionality constant
\eqn\hdg{
{\cal G}_{\rho t} ={\phi_e} \cdot \left({\cal G}_{\rho t}\right)_0~,\quad
{\cal B}_{t} ={\phi_e} \cdot \left({\cal B}_{t}\right)_0~.
}
The subindex ``0'' denotes the on-shell values we have used hitherto, corresponding 
to $\phi_e=1$. The $U(1)$ charge that generates gauge transformations is the 
variation of the on-shell action
\eqn\hdh{Q_{U(1)}= \left.{\delta S\over \delta \phi_e}\right|_{\phi_e=1}
= \int dt \sqrt{-\gamma}~J^a \left. 
{\delta{\cal B}_a\over\delta\phi_e}\right|_{\phi_e=1}
=\int dt \sqrt{-\gamma}\gamma^{tt} J_t {\cal B}_t
~.}
We used the definition of the current \hd. 

To make the formula explicit we insert the asymptotic behavior \hbc\ and
find 
\eqn\hdi{Q_{U(1)}=-{1\over \ell}\int dt\, e^{\psi }J_t~,}
to the leading order. This is a very conventional expression for the charge, essentially the time component
of the current.

As we have already remarked, the generator of diffeomorphisms \hdf\ will vanish, due to 
the overall metric factors. Therefore the path integral will be dominated
by the gauge theory generator \hdh. This is familiar from Sen's quantum entropy function. 
\refs{\SenVM\SenVZ}. 

\subsec{The Generator of Gauge Transformations}
We already showed that the generator of diffeomorphisms differs from the
the energy momentum tensor, due to large gauge fields near the boundary.  
Next we will show that the generator of gauge transformations differs from the
$U(1)$ charge, for the same reason. 

The variation of the action \heg\ under a gauge transformation 
$\delta {\cal B}_a = \partial_a \Lambda$ is
\eqn\wab{\delta_\Lambda S= \int dt  \sqrt{-\gamma}{\cal J}^a\partial_a\Lambda~,}
with
\eqn\wac{{\cal J}_t={\ell^3\over 2G_4}e^{-3\psi}{\cal B}_t~.}
To be more precise we proceed as for diffeomorphisms:
introduce a regulator function $F$ and define 
\eqn\wad{\delta_{\Lambda,F}\Phi\equiv\left( \delta_{F \Lambda}-F\delta_{\Lambda}\right)\Phi~.}
The variation of the action is then
\eqn\wbb{\eqalign{\delta_{F,\Lambda} S&= \int dt \sqrt{-\gamma}{\cal J}^a\Lambda\partial_aF\cr &=\sqrt{-\gamma}{\cal J}^t\Lambda - \int dt \, \partial_a\left(\sqrt{-\gamma}{\cal J}^a\Lambda\right) F~.}}
The last term in \wbb\ vanishes due to Noether theorem, and hence we identify the generator of gauge transformations as
\eqn\wbc{Q_\Lambda=\sqrt{-\gamma} {\cal J}^t = 
{\ell^3\over 2G_4} e^{-3\psi}\sqrt{-\gamma}\gamma^{tt}{\cal B}_t~.}
For example, we may evaluate this Noether charge on the black hole solution \hdb\ and find
\eqn\wbe{Q_\Lambda=-{\ell^2\over 2G_4}e^{-2\psi}\left(1-2{\epsilon\over\ell}e^{-\rho/\ell}+\cdots\right)~.
}
In the NHEK geometry \fgz\ the Noether charge takes the satisfying value
\eqn\wbf{
Q_\Lambda = - J~. 
}
The sign is just a matter of conventions. 

\subsec{Central Charge}

Now that we have identified the generators of diffeomorphism \hdf\ and gauge transformation \wbc, we can discuss if there is a central charge associated to the combined asymptotic symmetries constructed in section 5. 

The appropriate charge associated to the combined transformation $\epsilon^\mu$ \heb\ and $\Lambda$ \heg\ is a linear combination of $Q_{\epsilon}$ and $Q_{\Lambda}$. 
To properly add these two terms, we consider the asymptotic behavior of  the transformation parameters
\eqn\ra{\eqalign{\epsilon^t&=\xi(t) +2\ell^2 e^{-2\rho/\ell}\partial^2_t\xi+\ldots~,\cr
\Lambda&=  -2\ell\, e^{-\rho/\ell}\partial^2_t\xi+\ldots~.}}
To leading order we can write the gauge parameter as a function of the diffeomorphisms 
\eqn\ry{\Lambda=\ell\,{\cal B}_a\, \partial_\rho \epsilon^a+\ldots~.}
Therefore, when we considered the combined generator
\eqn\raa{Q_{(\epsilon+\Lambda)}(\epsilon+\Lambda)=Q_{\epsilon}\,\epsilon+Q_{\Lambda}\,\Lambda~,}
we will have a contribution from $Q_\epsilon$ given by \hdf\ and, from the gauge generator, a term of the form
\eqn\rya{Q_{\Lambda}\Lambda= - \sqrt{-\gamma}\gamma^{tt}{\cal B}_t {\cal J}_t\,\epsilon^t+\ldots~.}
where we used $Q_\Lambda{\cal B}_{t}\sim e^{\rho/\ell}$ to integrate by parts the $\rho$ derivative in \ry. 

This last expression has to be taken with some caution. From sections 6.1 and 6.3,  we have $Q_\Lambda\Lambda$ is of the same order as $Q_\epsilon \epsilon$, hence \rya\ is relevant to the transformation of \raa. But if we decompose \ra\ in Fourier modes, {\it i.e.} $\xi_n\sim e^{int}$,  the gauge function will only affect higher order modes ($Q_\Lambda$ will not contribute to the zero mode associated to the energy eigenvalue). When evaluating  \rya\ the transformation parameter $\epsilon^t$  has to be taken of order  $e^{-2\rho/\ell}$.

Lets first focus on the diffeomorphisms. From \hdf\ the relevant combination is $T_{tt}+{\cal B}_tJ_t$. Using \hej\ - \hex\ we find
\eqn\rab{\delta_{\epsilon+\Lambda}\left(T_{tt}+{\cal B}_tJ_t\right)=2\left(T_{tt}+{\cal B}_tJ_t\right)\partial_t \xi+\xi\partial_t\left(T_{tt}+{\cal B}_tJ_t\right)+{\cal O}(\sqrt{f}e^{-\rho/\ell})}
The generator transforms as a tensor of weight two as expected for a stress tensor of the CFT. The constant term in \hek\ does contribute a central term,  but it 
is cancelled by the contribution from gauge field. All the sub-leading corrections are proportional 
to the fluctuations $\sqrt{f}=\epsilon/\ell$ and redshifted, 
so asymptotically we have a vanishing central term for $Q_{\epsilon}$. 

The gauge contribution to the transformation of \raa\ reduces to
\eqn\ryb{\delta_{\epsilon+\Lambda}({\cal B}_{t}{\cal J}_t)=\xi\partial_t({\cal B}_{t}{\cal J}_t)-{\ell^3\over 2G_4}e^{-3\psi}\partial_t^3\xi}
%
%
%
As written, the weight of the tensor vanishes (corresponding to asymptotic behavior $e^{-2\rho/\ell}$) instead being two (corresponding to asymptotically constant). This is an artifact of the two
scales in the problem. As we already mentioned we are interested in 
$\epsilon^t\sim e^{-2\rho/\ell}$ and for such excitations the dimension is effectively two. 

Using the same normalization of central charge as in \hem, we identify the central charge as
\eqn\rae{c={6\over G_4}\ell^2_{AdS}~.}
This is the central charge associated to the charge $Q_{\epsilon+\Lambda}$. The diffeomorphism reflect the conformal structure of the generator, and after combined with the gauge transformation we obtain a central term. It is  also exactly what we expected from the discussion around equations \hem\ - \hep, but here we are able to compute the central charge from a  consistent generator.

\subsec{Level}
Finally, we consider the gauge transformation 
of the $U(1)$ current and the associated level. The level $k$ is defined as
\eqn\hew{\delta_\Lambda { J}_t={k\over 2}\ell_{\rm AdS}\partial_t\Lambda~.}
The gauge variation of the current \wac\ is
\eqn\heq{\delta_\Lambda { J}_t=e^{-2\psi}{\ell^2\over 2
G_4}\left(e^{-\psi}\ell\right)\partial_t\Lambda ~,}
so we find
\eqn\heqa{k={1\over  G_4}\ell^2_{\rm AdS}~.}
This value for the level is related to the central charge \rae\ in the simple manner 
\eqn\heu{c=6k~,}
as they were in our previous work \CastroMS. 

The relation \heu\ is precisely the one expected between central charge and the level 
of the R-current in $N=4$ supersymmetry. Since Kerr black holes break supersymmetry
completely, there would {\it a priori} be no reason to encounter this relation here. 
However, in the context of the larger picture we outlined in the introduction (and 
elaborate in the discussion), this result is precisely what we expect. According
to our interpretation the full theory carries $(4,0)$ supersymmetry and it is the
supersymmetric sector we probe here. That sector is in a {\it state} 
that breaks supersymmetry,
because of the condensate of quanta carrying the Kerr angular momentum, but
the {\it theory} should maintain supersymmetry. The relation \heu\ between central charge
and current gives support to this picture.

%
%

\newsec{Thermodynamics of Near-Extreme Kerr}
In this section we work out the consequences of the results in the
previous sections for the near extreme Kerr black hole. 

\subsec{The On-Shell Action}
We define AdS$_2$ quantum gravity as a Euclidean path integral (even though
we maintain Lorentzian notation), evaluated as function of boundary data. In 
particular it is a function of the inverse temperature and the Noether
charge. 

The Euclidean version of the black hole 
metric \hdb\ has a conical singularity unless the Euclidean time coordinate has 
periodicity
\eqn\ma{
\beta_0 = {2\pi\ell\over\sqrt{f}} ={2\pi\ell^2\over\epsilon} ~.
}
In asymptotically AdS$_2$ the physical inverse
temperature becomes
\eqn\maa{
T^{-1} = \beta= {2\pi\ell^2\over\epsilon}\cdot {1\over 2}e^{-\psi}e^{\rho/\ell} (1 - {\epsilon^2\over\ell^2}e^{-2\rho/\ell})~.
}
In boundary formulae like this one $\rho$ is interpreted as a cut-off, as usual. In the
limit $\rho\to\infty$ where the cutoff is removed, the physical temperature $T\to 0$,
as expected for extremal black holes. We want to keep the finite temperature
effects near extremality so we will keep the leading corrections, rather than taking
the strict limit.   

We evaluate the path integral in a sector where the Noether generator \wbe\ takes
a fixed value which we identify with the angular momentum or the central charge
\eqn\mba{
Q_\Lambda=-{\ell^2\over 2G_4}e^{-2\psi}\left(1-2{\epsilon\over\ell}e^{-\rho/\ell}+\cdots\right)
= - J = - {c\over 12}~.
}

We can now compute the on-shell action from our renormalized 
boundary action \hcg. The bulk terms in the action \hcg\ vanishes on the 
solution \hdb, as one can readily show using the equations of motion \fgd, \fge. 
The boundary terms give the on-shell action
\eqn\mab{\eqalign{
I_M 
&= {\ell^2\over 2G_4} \int dt \sqrt{-\gamma} 
[ e^{-2\psi}K - {1\over 2\ell} e^{-\psi} + {\ell\over 2} e^{-3\psi} {\cal B}_a {\cal B}^a] \cr
&= {\ell^2\over 2G_4}\cdot \beta\cdot {2\epsilon\over\ell^2} e^{-\psi}e^{-\rho/\ell} [ 1 - {\epsilon\over\ell} e^{-\rho/\ell} + \ldots]\cr
& = {\pi\ell^2\over G_4} e^{-2\psi} [ 1 - {\epsilon\over\ell} e^{-\rho/\ell} + \ldots]\cr
& = {\pi c\over 6}  [ 1 +  {\epsilon\over\ell} e^{-\rho/\ell} + \ldots] \cr
&= {\pi c\over 6} (1 + \pi T\ell_{\rm AdS})~ ~.
}}
We made the final expression more transparent by introducing the
central charge using \mba\ and then the temperature \maa. 

The Euclidean action (which has the opposite sign of the Lorentzian action \mab)
is essentially the free energy
\eqn\mad{
\beta F = I_E =  - I_M = - {\pi c\over 6} (1 + \pi T\ell_{\rm AdS})~.
}
The entropy derived from the renormalized on-shell action then becomes
\eqn\mae{
S = - {\partial F\over\partial T}   =  {\pi c\over 6} + {\pi^2c\over 3} T\ell_{\rm AdS} ~.
}
The first term can be interpreted as the ground state entropy of the chiral half of
a CFT with central charge $c_L=c$, as in the usual Kerr/CFT \GuicaMU. The 
second term is the Cardy expression for the other chiral half of the CFT with the 
same central charge $c_R=c$, excited to temperature $T$. The entropy \mae\ computed 
from the renormalized on-shell action is equivalent in form to the 
expression \je\ we found directly from the standard Kerr entropy. 

\subsec{The Excitation Energy}

Our computation of the on-shell action assigns the energy
\eqn\maf{
E = {\partial (\beta F)\over\partial\beta} = {\pi^2 c\over 6} T^2\ell_{\rm AdS}~,
}
to the excitations of the CFT. As a check on our computations we would like
to recover this expression directly from the energy momentum tensor. 

The energy is generated by time translations, {\it i.e.} diffeomorphisms with $\epsilon^t=1$.
In this case the compensating gauge transformations \heg\ vanish the energy can
be computed directly from the (shifted) energy momentum tensor \hdfa. We find
the local energy
\eqn\mag{
E_{\rm loc} = T_{{\hat t}{\hat t}} + {\cal B}_{\hat t} J_{\hat t}
= {\epsilon^2\over \ell G_4} e^{-\psi}e^{-2\rho/\ell}
= {\pi^2\ell^3\over G_4} T^2 e^{-3\psi} = {\pi^2 c\over 6} T^2\ell_{\rm AdS}
~.
}
We used the boundary metric \hbf\ to go to the local frame and then traded
the cutoff for the temperature using \maa. In the final step we used the central charge
\rae. Our final expression \mag\ computed from the local energy momentum
tensor agrees with \maf\ computed from the on-shell action. 

In our ``phenomenological'' analysis of the Kerr entropy, we detemined the CFT level 
$h_R$ as \jj, using the macroscopic area law. In particular we found that the excitation
energy should be of second order in the decoupling parameter $\lambda$, as expected
for excitations described in the frame rotating at (nearly) the speed of light. In the CFT
we similarly assign the excitations an energy \maf\ of second order in the small 
parameter $T$ or, equivalently, or second order in the cut-off $e^{-\rho/\ell}$. The implied
correspondence between parameters is
\eqn\mah{
\lambda^2 = 2e^{-2\rho/\ell}~.
}
It would be interesting to have an independent derivation of the relative numerical factor. 

\newsec{Discussion}
There are by now several approaches to Kerr/CFT and, more generally, to AdS$_2$ quantum
gravity. The subject faces many challenges and its entire consistency has been challenged. 
A strategy that may go far towards addressing the challenges that remain would be to 
relate the various approaches more precisely. In this spirit we comment on how our work
is related to some other ideas: 

\medskip
\noindent
The {\bf Quantum Entropy Function} is a rather systematic approach to Euclidean 
AdS$_2$ quantum gravity \refs{\SenVM,\SenVZ}. It is well established and it satisfies many 
consistency checks. Our approach reduces to the quantum entropy function in the extremal
limit and then generalizes to where small excitations above extremality are included. 
To make this generalization explicit in the quantum entropy framework it must be shown
that boundary conditions can be deformed consistently.

\medskip
\noindent
The {\bf 3D origin of AdS$_2$ Quantum Gravity} could help clarify many issues. The Euclidean
version of this is quite well understood \GuptaKI\ but the Lorentzian theory raises some
issues \refs{\BalasubramanianBG}. In Lorentzian signature AdS$_2$ 
can be embedded in to AdS$_3$ 
along a light cone direction. It then seems that our 2D theory can be interpreted as a 
slightly deformed embedding, along a direction that is near the light-cone but does not 
coincide with it. This would explain why we have access to both chiralities of the theory, 
corresponding to the direction along the light cone and transverse to it. 

\medskip
\noindent
The {\bf Asymptotic Symmetry Group of Kerr} involves the theory directly in 4D. We have
reduced to 2D in a straightforward manner but the consistency was not established in detail. 
It would be interesting to extend our approach to the full NHEK geometry. That could help
address some of the challenges facing the original Kerr/CFT \refs{\AmselEV,\DiasEX,\AmselPU}, 
by providing a flexible Lagrangean framework. In addition this might provide some insight to  boundary conditions and asymptotic symmetries studied in \refs{\MatsuoSJ,\MatsuoPG,\RasmussenIX}. 

\medskip

The justified focus on the immediate challenges to Kerr/CFT should not distract from 
the larger opportunities that seem to be appearing. The important point is not just 
whether or not the existence of 2D conformal symmetry and the applicability of 
Cardy's formula can be established on completely general grounds. The apparent 
ubiquity of Virasoro algebras is a surprise either way. In physical terms, 
many black holes apparently have an ``effective string'' structure in the spacetime 
theory, with excitations classified according to two chiral sectors of a CFT. It is far from 
obvious {\it a priori} that such a structure should apply to extremal Kerr, especially since the
isometry group is just $SL(2,\RR)\times U(1)$. It begs the question of whether more generic black holes, 
without any $SL(2,\RR)$ isometry at all, would similarly have two Virasoro algebras.

There is a hint of such a structure in the general entropy formula for Kerr 
\eqn\ka{\eqalign{
S & ={A\over 4G_4} =  2\pi \left(M^2 l^2_P + \sqrt{M^4 l^4_P - J^2}\right)~,
}}
which might be interpreted as a Cardy formula that applies all the way off 
extremality \LarsenBU. 
Indeed, the entropy of much more general 4D black holes (with many charges) 
takes a very similar form, with the angular momentum $J$ appearing as
a subtraction under the square root on the ``supersymmetric'' (R) side \CveticXV. 
The two apparent levels satisfy a ``level matching condition" \LarsenGE
\eqn\kb{
N_L - N_R = (M^4 l^4_P)  - (M^4 l^4_P -J^2) = J^2~,
}
that is integral also in much more general cases. We referred to this level matching
condition already in the introduction. The general level matching 
condition can be expressed geometrically in terms of the outer and inner horizon areas as
\refs{\CveticUW,\CveticXV}
\eqn\kc{
{1\over (8\pi G_4)^2}A_+ A_- ={\rm integer}~.
}
The origin of this type of relation is far from clear. It seems to indicate that the
geometry ``knows'' about the division into two chiralities quite generally and that,
in some way, the general theory is a combination of two factors, each of which has
the structure familiar from BPS theory. This kind of prospect is perhaps the most
compelling motivation for further development of the field. 

\bigskip
\noindent {\bf Acknowledgments:} \medskip \noindent

We are grateful to A. Strominger for discussion and encouragement. 
We also thank V. Balasubramanian, A. Dabholkar, C. Keeler, A. Sen, and J. Simon for discussion. 
The work of AC is supported in part by the National Science and Engineering Research 
Council of Canada. The work FL is supported in part by US DoE.
FL also thanks the Aspen Center for Physics for hospitality while this work was completed.

\listrefs

\end